\documentclass[fleqn,usenatbib]{mnras}

\usepackage{mathptmx}
\usepackage{txfonts}

\usepackage[T1]{fontenc}

\DeclareRobustCommand{\VAN}[3]{#2}
\let\VANthebibliography\thebibliography
\def\thebibliography{\DeclareRobustCommand{\VAN}[3]{##3}\VANthebibliography}

\usepackage{graphicx}	
\usepackage{amssymb}	

\title[Coronal flow speeds : Results from Akatsuki]{Turbulence dynamics and flow speeds in the inner solar corona: Results from radio-sounding experiments by the Akatsuki spacecraft}

\author[Jain et al.]{
Richa N. Jain $^{1,2}$ \thanks{E-mail: richanajajain@gmail.com},
R. K. Choudhary $^{1}$,
Anil Bhardwaj $^{3}$,
T. Imamura $^{4}$,
Anshuman Sharma $^{5}$, Umang M. Parikh $^{5}$
\\
$^{1}$ Space Physics Laboratory, VSSC, ISRO, Thiruvananthapuram - 695022, Kerala, India \\
$^{2}$ Research Center, University of Kerala, Thiruvananthapuram, - 695034 Kerala, India\\
$^{3}$ Physical Research Laboratory, Ahmedabad -380009, India\\
$^{4}$ Graduate School of Frontier Sciences, The University of Tokyo, Kiban‐tou
4H7, 5‐1‐5 Kashiwanoha, Kashiwa, Chiba 277‐8561, Japan \\
$^{5}$ ISRO Telemetry, Tracking and Command Network (ISTRAC), Bengaluru - 560058, India}

\date{Accepted 2023 August 14. Received 2023 July 18; in original form 2023 February 07}

\pubyear{2023}

\begin{document}
\label{firstpage}
\pagerange{\pageref{firstpage}--\pageref{lastpage}}
\maketitle

\begin{abstract}

The solar inner corona is a region that plays a critical role in energizing the solar wind and propelling it to supersonic and supra-Alfvenic velocities. Despite its importance,  this region remains poorly understood because of being least explored due to observational limitations. The coronal radio sounding technique in this context becomes useful as it helps in providing information in parts of this least explored region. To shed light on the dynamics of the solar wind in the inner corona, we conducted a study using data obtained from coronal radio-sounding experiments carried out by the Akatsuki spacecraft during the 2021 Venus-solar conjunction event. By analyzing X-band radio signals recorded at two ground stations (IDSN in Bangalore and UDSC in Japan), we investigated plasma turbulence characteristics and estimated flow speed measurements based on isotropic quasi-static turbulence models. Our analysis revealed that the speed of the solar wind in the inner corona (at heliocentric distances from 5 to 13 solar radii), ranging from 220-550 km/sec, was higher than the expected average flow speeds in this region. By integrating our radio-sounding results with EUV images of the solar disk, we gained a unique perspective on the properties and energization of high-velocity plasma streams originating from coronal holes. We tracked the evolution of fast solar wind streams emanating from an extended coronal hole as they propagated to increasing heliocentric distances. Our study provides unique insights into the least-explored inner coronal region by corroborating radio sounding results with EUV observations of the corona.

\end{abstract}




\section{Introduction}

The unusually high temperature of the solar corona doesn't let it end at a sharp boundary; instead, it continues outward to the interplanetary medium. This thermally driven flow of ionized particles is termed the solar wind \citep{Parker1958}. Observations made from space confirmed that this ubiquitous flow of charged particles in the form of solar wind pervades the entire interplanetary medium at all the heliolatitudes and heliolongitudes \citep{McComas2000}. These observations also revealed the fact that the specific regions in Corona give rise to various kinds of solar wind streams. Solar wind observations near 1 AU have identified three main types of solar wind based on differences in their speeds and densities: the fast solar wind, with speeds greater than 400-700 km/s, the slow solar wind, with speeds around 400 km/s, and transient wind that occurs during coronal mass ejections \citep{Marsch1999}. These different types of wind are thought to originate from different regions of the corona. For example, fast solar wind typically originates in coronal holes, which are darker regions (as viewed in the EUV and soft X-ray coronal images) of the Sun at higher heliolatitudes with open magnetic field lines \citep{Zirker1977}. Slow solar wind, on the other hand, originates above the active regions in the streamer belts, which are mostly found near the low-equatorial heliolatitudes \citep{Geiss1995}. The characteristics and extent of these coronal features are determined by the convective activity of magnetized plasma that surfaces from the photosphere beneath the corona. The relative motion and abundance of these features depend on various factors, including the phase of the solar activity cycle, the number of sunspots, and the frequency of magnetic reconnection processes occurring on the solar surface \citep{McComas2003, Antiochos_2011}.

It is also observed that Solar wind plasma velocities beyond 20-30 $R_\odot$ (1 $R_\odot$ = 1 solar radii = 696,700 Km) and near-Earth regions are mostly supersonic and super-Alfvenic, with average magnitudes of between 400 to 700 km/s, corresponding to slow and fast wind respectively. The solar wind in the inner corona, on the other hand, is subsonic \citep{Woo1995, Sheeley1997}. The acceleration of solar wind particles from subsonic to supersonic and super-Alfven velocities occur in the transition region \citep{Kasper2021, Mcintosh2011}.  Thus, mapping solar wind speeds near the Sun becomes crucial to understanding its origin and expansion.

According to Parker's hydrodynamic Corona theory, the high temperature of the Corona is the main factor responsible for accelerating the solar wind by creating a pressure gradient that propels ionized charged particles outward from the corona and toward the interplanetary medium \citep{Parker1958}. However, the physics behind coronal heating mechanisms, solar wind acceleration processes, and phenomena occurring at the transition of the solar wind at the critical Alfvénic surface \citep{Adhikari2019} still require deeper understanding. Various mechanisms have been proposed, with the most prominent attributing coronal heating to turbulence and the dissipation of energy contained in magnetohydrodynamic (MHD) waves that traverse the low-beta plasma region \citep{Cranmer2015}. Some models also consider small-scale impulsive energy releases by magnetic reconnection during nano-flare events \citep{Sakurai2017}. Recently, the XSM instrument onboard Chandrayaan-2 has detected a large number of micro-flares (sub-A class flares) during the deepest minima of the last century \citep{Vadawale2021}. Estimates on nano-flare heating of the corona \citep{Upendran2022} have also been made, providing further evidence for the nano-flare mechanism.
 
There are several ways in which MHD waves mediate turbulence, such as interaction of forward propagating Alfvenic waves with counter-propagating waves, partial reflection due to coronal inhomogeneity, mode coupling/phase mixing in regions of strong gradients of density and magnetic field, and especially for low beta plasmas parametric decay instability (PDI) mechanism, leading to the to the energization of the solar wind in the solar Corona. These mechanisms have a rich scientific basis and many detailed studies are available \citep{Hollweg1986, Matthaeus1999, Cranmer2015, Shoda2019}.  Among them the conversion of Alfven waves into compressed ion-acoustic waves and their dissipation of shock waves emerges as significant one. The Alfven waves, which are transverse waves, carry photospheric magnetic field into the base of the corona and the transition region. The energy fluxes transported by the Alfven waves are sufficient to energize the quiet corona and power the fast solar wind \citep{Mcintosh2011}. Non-linear processes in the turbulent plasma cause Aflvenic waves to decompose into forward propagating compressive magnetosonic waves and backward propagating alfvenic waves \citep{Shoda2019}. The interactions between the outgoing and incoming wave-packets sets up the turbulence in the medium. Turbulence, in turn, assists in the energy cascading process and magnetosonic waves dampen and dissipate energy leading to high temperatures in Corona \citep{Coleman1968, Cranmer2007}, and also causes a wave pressure gradient that accelerates the wind \citep{Alazraki1971, Belcher1971}. Several models based on simulation of MHD waves turbulence and dissipation \citep{Roberts1992, Verdini2007, Cranmer2007, Shoda2019, Chandran2019} have been proposed to explain this phenomenon of plasma turbulence, energy dissipation and acceleration of solar wind.

\begin{figure}
    \centering
    \resizebox{\hsize}{!}{\includegraphics{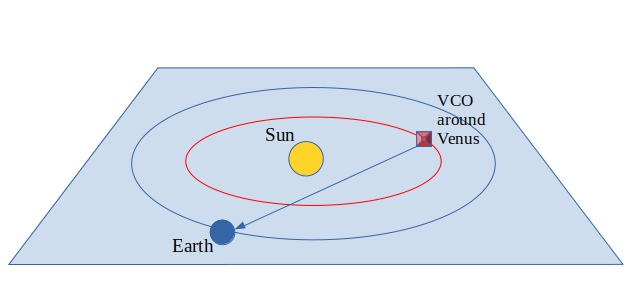}}
     \caption{Schematic of Venus-Sun-Earth during 2021 Venus-Solar Conjunction in the Ecliptic plane. VCO (Akatsuki) spacecraft is orbiting around Venus.}
     \label{SVE}
\end{figure}

The harsh conditions in the solar corona, including high temperatures and a complex magnetic field, have made in-situ measurements challenging. In this context, coronal radio sounding techniques have proven invaluable for obtaining information about the near Sun coronal medium, where direct plasma measurements are difficult. Early studies used radio emissions from natural sources like the Crab Nebula and quasars to investigate Interplanetary Scintillations (IPS) at different distances from the Sun \citep{Hewish1964}. Subsequently, it was realized that higher frequency and more stable radio beacons, such as spacecraft signals during radio occultation in the S-, X-, and Ka-bands, can be utilized to probe regions of the inner corona with higher electron densities \citep{Bird1992}. Previous missions have utilized downlink radio signals from spacecraft during solar conjunction periods to conduct studies on the solar corona \citep{Wexler2019, efimov2005A, Imamura2005, patzold1997, Chashei2005, Efimov2008, Patzold2012}. Solar conjunction refers to the period when a planet (and hence the spacecraft orbiting that planet), the Sun, and the Earth align in a straight line, with the Sun at the center. Radio signals transmitted from the spacecraft pass through the solar corona during this period (Figure \ref{SVE}). As the corona is an ionized medium with fluctuating electron density, it induces dispersive signatures in the radio signals. These signatures can be analyzed to study various coronal properties, including plasma turbulence regimes, velocity, density, magnetic field fluctuations, and solar wind acceleration \citep{WooArmstrong1979, Bird1982, BirdEidenhofer1990, Wohlmuth2001, efimov2005}.

In the present study, we used the single downlink X-band radio signals transmitted by the Akatsuki spacecraft (also known as Venus Climate Orbiter) to study coronal properties during Venus-Solar Conjunction that occurred in March-April 2021. This is the period that corresponds to the ascending phase of solar cycle 25 when the solar activity started increasing after minima of comparatively weaker solar cycle 24. A serendipitous observation has been the fast solar wind during the period of study which gave us an opportunity to track the evolution of the fast solar wind speed from the source to different heliocentric distances to the L1 point in the near-Earth environment as observed during the experiment period. The radio-sounding data were collected simultaneously at two stations - IDSN (Indian Deep Space Network), Bangalore, and at UDSC (Usuda Deep Space Center), Japan. Here, we have studied the plasma turbulence characteristics and estimated the flow speeds at the heliocentric regions as close as 5-13 $R_\odot$ using radio signals received at IDSN Bangalore as discussed in further sections.

\section{Data processing and Observations}

\begin{figure*}
     \centering
     \centerline{\includegraphics[draft=false,width=1.15\textwidth]{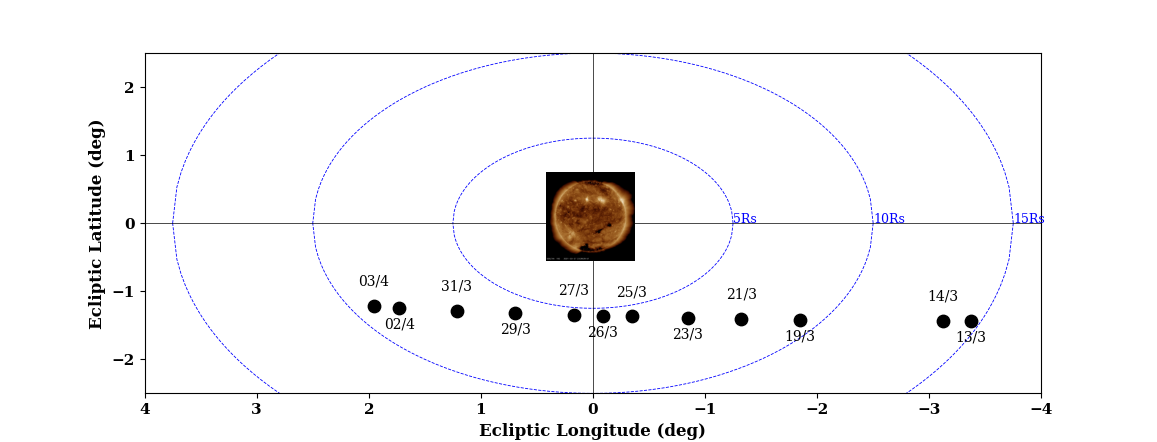}}
     \caption{ Akatsuki-solar conjunction geometry during March-April 2021. Circles shows apparent position of Akatsuki (in DD-MM format) as seen from plane of sky on Earth with respect to the Sun at 00:00 UTC on experiment days in Solar Ecliptic (ECLIPJ2000) coordinate system. SDO/AIA image of coronal disc on 2021 March 27 is embedded depicting Sun. Concentric circles are constant heliocentric distance contours drawn at every 5 $R_\odot$.}
     \label{ROgeo}
\end{figure*}

The experiment was performed using the Akatsuki orbiter which is a Japanese satellite orbiting Venus \citep{Nakamura2011}. During the 2021 Venus-Solar Conjunction, the solar occultation experiments were performed using the Radio Science payload onboard Akatsuki \citep{Imamura2011}. It uses an onboard ultra-stable oscillator (USO) to generate radio signals at X-band (8410 MHz) frequencies. Experiments are conducted in one-way downlink mode and signals are recorded at IDSN, and UDSC, simultaneously or individually depending upon the satellite's orbit geometry vis-a-vis station visibility. Further details of the payload, technical specifications, and data analysis methods are given elsewhere \citep {Imamura2011, Imamura2017, Tripathi2022}. This conjunction geometry is important in two aspects: first, the solar proximate distance, i.e. tangential distance of the closest point of radio ray path from the center of the Sun, is as close as 4-13 $R_\odot$ and crosses the heliolatitudes between 24 - 86$^\circ$ S. Such a geometry provided us with a good opportunity to probe the spatial variations in plasma conditions near southern solar hemisphere corona. Figure \ref{ROgeo} depicts the relative position of spacecraft and the Sun as seen from Earth's sky plane. Second, this experiment was performed during March 2021, a period that corresponds to the ascending phase of solar cycle 25 (solar cycle 24 minima was in 2019). During our observation period, solar 10.7 cm flux was 76.5 SFU (1SFU = $10^{-22}$ Watts per meter square per Hertz) and the average sunspot number was 17.2 (\url{https://www.swpc.noaa.gov/products/solar-cycle-progression}). No strong CME event was recorded during this period, thus no transient wind existed and this makes it an ideal situation to study the background solar winds in quiet solar conditions.

The radio signal data were recorded simultaneously at the IDSN, Bangalore, and UDSC, Japan using open-loop receiver systems in the standard CCSDS - RDEF (Raw Data Exchange Format) format (\url{https://public.ccsds.org/Pubs/506x1b1.pdf}). The data collected at these stations were later used to calculate the experimentally recorded observed Doppler frequency fluctuations imprinted on radio signals passing through the solar corona, by adopting standard processing methods as given in \citep {Patzold2004, Choudhary2016, Tripathi2022}. The observed Doppler was extracted at the sample rate of 1 Hz. For each station, theoretical Doppler (range rate Doppler, or Doppler arising due to relative motion of spacecraft and receiver) was calculated using SPICE kernels provided by UDSC/IDSN teams \citep{SPICEcite}. The difference between observed and theoretical Doppler gives an observable quantity of ``differential Doppler-frequency residual'' values, that separates the non-dispersive Doppler from the dispersive effects \citep{Tripathi2022}. The appropriate fit was calculated and subtracted from processed Doppler-frequency residual values to compensate standard sources of error for the oscillator drifts, other imperfections in trajectory, and the effects of the terrestrial ionosphere and atmosphere \citep{Tripathi-Choudhary2022}. We obtained fluctuations of these frequency residual values around zero mean termed frequency fluctuations (FF), a parameter that contains the dispersive effects arising due to the turbulence in the propagating medium. The proximate distance of radio ray path from the solar center (termed solar offset) in terms of heliocentric distances is calculated using the ephemeris of the Akatsuki spacecraft and planetary ephemeris as offered in NASA SPICE tool \citep{SPICEcite}.  The results of the turbulence regimes have been quite successfully used by the radio-science community to understand various phenomena such as electron density fluctuation spectra \citep{WooArmstrong1979}, phase scintillation spectra \citep{Patzold1996}, temporal spectra of fluctuations amplitude of radio signals \citep{Yakovlev1980} and magnetic field power spectra \citep{Chashei2000, Wexler2019b} using recorded faraday rotation in the received signal from solar conjunction experiments. In this present work, we used these FF values to characterize the turbulence spectrum spectra and estimate the plasma flow speeds in the inner coronal region using the methodologies as prescribed in the radio-sounding data analysis \citep{Wexler2019}.

\section{Results}

As mentioned, in this study we utilized the observable parameter obtained from our radio occultation experiments, the Doppler frequency fluctuations (FF). Fast Fourier transform was applied to these FF. The resultant temporal spectrum of Fourier transformed values of FF was approximated as the single power-law spectrum between frequency range [$\nu_{low}$, $\nu_{up}$]. The frequency range to apply power-law approximation is defined such that the lower frequency cutoff $\nu_{lo} = 10^{-3} Hz$ is the length of the processed interval and upper $\nu_{up} = 5 \times 10^{-2} Hz$ is bounded by one-tenth of Nyquist frequency under consideration of the fact that above which system noises will dominate the fluctuation data \citep{Efimov2008}. The linear fit on the log-log plot gives a slope that is an exponential index ($\alpha_f$) of the single-power law spectrum of recorded frequency fluctuations. $\alpha_f$ is related to the spectral index of electron density turbulence spectrum p as $\alpha_f = p-3$ for 3D turbulence \citep{Jain2022, Efimov2008}.

Table \ref{Alphatable}, shows the values of the subsequent spectral index for all the experimental days. It can be observed that spectrum has slope values ranging from 0.12-0.30 for the entire duration of conjunction experiment.

\begin{table}
\centering
\begin{tabular}{|c|c|c|}
\hline
Date & R($R_\odot$) & Spectral Index ($\alpha_f$) \\
\hline
14-March-2021 & 12.82 & 0.24 \\
\hline
19-March-2021 & 8.71 & 0.29 \\
\hline
21-March-2021 & 7.24 & 0.16 \\ 
\hline
23-March-2021 & 6.09 & 0.12 \\
\hline
25-March-2021 & 5.28 & 0.21 \\
\hline
26-March-2021 & 5.09 & 0.23 \\
\hline
27-March-2021 & 5.08 & 0.20 \\
\hline
29-March-2021 & 5.57 & 0.18   \\
\hline
31-March-2021 & 6.62 & 0.28 \\
\hline
02-April-2021 & 7.98 & 0.13 \\
\hline
\end{tabular}
\vspace*{.25cm}
\caption{Index of temporal spectrum of frequency fluctuations recorded at IDSN station obtained from Akatsuki Solar-Conjunction 2021 experiment. Values of heliocentric distance $R_\odot$ of ray path proximate point and spectral Index corresponding to each experiment day are tabulated. }
\label{Alphatable}
\end{table}

The theory of solar wind acceleration by the MHD waves and turbulence states that the turbulence is caused by the ensemble of Alfven and magnetosonic waves entrained in the solar wind. The transition from the acceleration region to the steady-flow region is accompanied by the spatial change in turbulence regime \citep{ChasheiModel1983}, as reflected by the changes in the slope of the turbulence spectrum. It is known that for the inner coronal regions, the spectral index values are generally lower and appear as flatter slopes in the turbulence spectrum. Lower $\alpha_f$ ($\sim0.2-0.3$) values in temporal frequency fluctuation spectra denote the energy source/input regime where turbulence is still underdeveloped \citep{Armand1987}.  With the increasing heliocentric distances, the cascading of energy sets up, and thus the slope value starts increasing. The spectral index of a fully developed three-dimensional Kolmogorov electron density turbulence spectrum is $p = 11/3 $  \citep{Woo1976}. This corresponds to the spectral index $\alpha_f\sim 0.6 - 0.7$ of frequency fluctuation spectrum which denotes an inertial range of developed turbulence regime and occurs generally around the transition region beyond 10 $R_\odot$ \citep{Efimov2008}. Previous solar occultation studies from spacecraft \citep{efimov2005, Jain2022} have shown that as we move away from the solar surface, with the increasing heliocentric distances the slope of plasma turbulence spectrum gets steeper and transition to the Kolmogorov values usually occur around 10 - 15 $R_\odot$ \citep{Imamura2005}.

\begin{figure*}
    \centering
    \resizebox{\hsize}{!}{\includegraphics{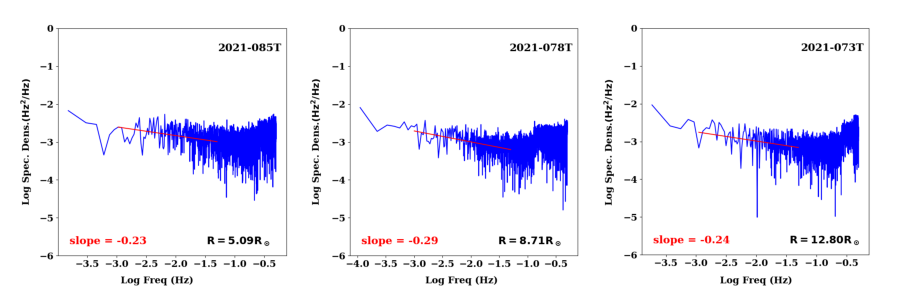}}
   
     \caption{Temporal Spectrum of frequency fluctuations (FF) on three experiment days corrosponding to the increasing heliocentric distance of proximate ray path point (a) When R = 5.09 $R_\odot$ the index is 0.23 (b)When R = 8.71 $R_\odot$ the index is 0.29 (c)When R = 12.80 $R_\odot$ the index is 0.24 . The flatter slope values even with increasing heliocentric spectrum indicates the radio ray path was embedded in the region of fast solar wind. }
     \label{FFSpec}%
     \end{figure*}

However, in our case, we found that the turbulence spectrum was flatter for the entire range of $5-13 R_\odot$ heliocentric distances. Figure \ref{FFSpec} shows the turbulence spectrum for three such experiment days where each day represents the different solar proximate distance. It can be noted that at closer heliocentric distances (first spectrum when $R = 5.09 R_\odot$ the slope is 0.23), the spectrum is flatter. With the increasing heliocentric distances it is expected that slope values increase which should manifest as steeper spectrums. However, for the other two spectra in Figure \ref{FFSpec} (when $R=8.71 R_\odot$ the slope is 0.29 and when $R=12.80 R_\odot$ the slope is 0.24) it is observed that with increasing heliocentric distances the turbulence spectrum remains still flatter.

The intriguing nature of this observation prompted us to investigate the existing coronal dynamics conditions prevailing during the experimental days with respect to heliolatitudes and coronal features of the proximate point of the radio ray path. Previous studies on the turbulence characteristics of inner solar wind plasma, which observed the dependence of spectral index with respect to the heliolatitudes, have reported that at the higher heliolatidues, the spectral index values are generally lower \citep{Patzold1996, Efimov2010}. During an ascending phase of the solar cycle, when the solar activity is lower, higher heliolatitudes are dominated by the features known as coronal holes, which are sources of fast solar winds. The fast solar winds emanate from the coronal holes which are regions of open field lines. This leads to the lower values of the sound waves to Aflvenic wave speed ratio and thus turbulence takes a longer time (hence larger heliocentric distances) to achieve a fully developed spectrum. Hence, previous studies \citep{Patzold1996, Efimov2010} have shown that at higher heliolatitudes the turbulence spectrum is expected to still be an underdeveloped regime and is indicated by the flatter slopes. The change in the turbulence regime takes place at farther greater distances from the solar surface for the fast solar wind than for the slow solar wind.

To put the discussions in context, we estimated the solar wind flow speeds from our spectral index calculations. The subsequent section discusses the estimation of solar wind flow speeds using a well-known formulation, known as the isotropic quasi-static turbulence model, which is based on the propagation of electromagnetic waves through a turbulent plasma medium and is readily employed in studying the turbulence characteristics of inner solar wind \citep{Armand1987, ChasheiModel1983}. It is worth mentioning here that the solar wind speed can be estimated from the intensity scintillation spectrum as well, as done by \citep{Chiba2022}. The results will be reported elsewhere.

\section{Discussions}

Our study shows that the turbulence spectra for the given heliocentric distances of solar offsets are generally flatter. This is interesting because there should have been a gradual increase in the slope values with the increasing heliocentric distances \citep{Jain2022}. As a flatter slope indicates the source region in the turbulence regime, we need to understand why the solar wind remains in the source region of energization and how the solar wind during this period evolves with varying heliocentric distances.

Turbulence is a pervasive characteristic of heliospheric plasma across all heliocentric distances and latitudes \citep{efimov2005A}, and turbulence models employed in the study of solar wind plasma successfully capture the energy transfer in Magnetohydrodynamic (MHD) waves, the heating of the solar corona, and the outward acceleration of the solar wind. At closer heliocentric distances near the active source regions, Aflven waves emerge from the base of the solar corona carrying large amounts of energy. As they propagate forward, the magnetic field energy decreases, and the non-linear processes such as parametric decay instability activates in the low-beta plasma. Then a forward propagating Alfven wave is converted into a backward propagating MHD wave and forward propagating ion-acoustic (magnetosonic both compressive and non-compressive modes) waves.  These magnetosonic waves propagate longitudinally to create random density and magnetic fluctuations in plasma, thereby further setting up a turbulence cascade in the medium. The energy contained in magnetosonic waves is then transferred into the lower wavelengths which dissipates and heat up the corona, leading to the acceleration of solar wind particles \citep{Coleman1968, Goldstein1978, Derby1978, ChasheiModel1983, Jain2022}.

\subsection{Velocity estimation from two-station cross correlations and its limitations}

As mentioned earlier, the signal transmitted from the Akatsuki spacecraft during the coronal-sounding experiment was received simultaneously at two stations - UDSC (Japan) and IDSN (India). This encouraged us to use one subclass of these coronal-sounding experiments, the VLBI (Very Long Baseline Interferometer) technique to estimate flow speeds \citep{Spangler2002}. In this technique, the radio signal transmitted by the spacecraft is received simultaneously at two spatially separated ground stations. The radio signals received at two antennas are then correlated to deduce the observable difference in the Doppler phase of two signals received, which can be attributed to the turbulent plasma irregularities that are embedded in the solar wind and advected radially across the line of sight of two antennas. The frequency residuals and cross-correlation functions are calculated in overlapping time intervals of the same experiment day for both stations. It is assumed that the density homogeneity of the medium, which causes frequency fluctuation in the received radio signal, is a ‘frozen-in’ solar plasma medium and is convected outwards in a direction perpendicular to the signal ray path with solar wind speed. The coronal separation of ray path proximate points $\Delta R$  at the time when signals are received at two stations, combined with a correlation time lag $\tau$, assists in estimating the average outward flow speeds ($V_{sw} = \Delta R / \tau $ ) of inner solar wind \citep{Janardhan1999, Efimov2010, MaoliMa2021}.

\begin{figure}
    \centering
    \resizebox{\hsize}{!}{\includegraphics{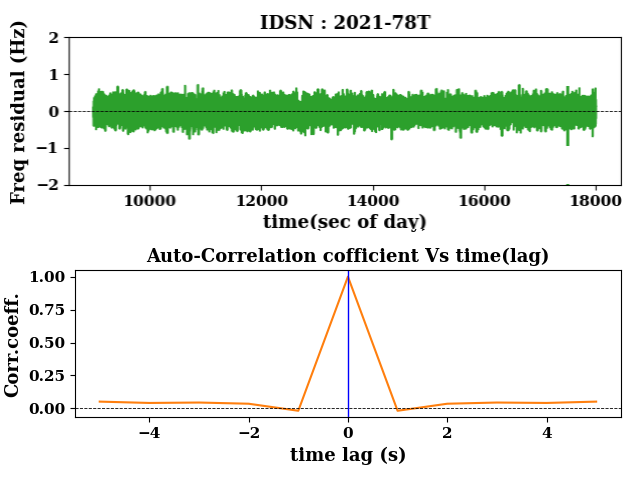}}
 
     \caption{Upper Panel -  Frequency fluctuation residual values from the radio signal obtained from IDSN station. Lower Panel - Autocorrelation of doppler frequency residuals obtained from radio signal received at IDSN station on March 19, 2021 from the Akatsuki coronal sounding experiment.}
     \label{autocorr}%
     \end{figure}

We used a similar approach, with our Doppler residual data. It was however found that the cross-correlation between the two station data was consistently weak and correlation coefficient values were poor. The weak correlation made the determination of correlation time lag values uncertain and hence we were not able to use this method to calculate the flow speeds conclusively. One of the reasons for this weak cross-correlation in two data sets can be the absence of this density inhomogeneities markers across the radio ray path in the region of the coronal medium which is being probed \citep{Janardhan1999}. To support this observation, we considered the auto-correlation of the single station data and found out that the correlation coherence length, which is indicative of the spatial span and lifetime of the discernible inhomogeneity plasma tracer to travel across separation between two radio-ray paths, was actually too narrow to estimate flow speeds. For instance Figure \ref{autocorr} represents the autocorrelation of the Doppler frequency fluctuations recorded at the IDSN station on experiment day of March 19, 2021, when the area probed by radio signals is around $8.7 R_\odot$. During this day the average coronal separation of the two radio paths one at IDSN and the other at UDSC station was $\Delta R$ = 2636 km. The auto-correlation of frequency residual values at IDSN station yields the coherence length of $\Delta \tau$ < 1 sec. If we assume that the average solar wind speed at this coronal region is approximately 200 km/s \citep{Sheeley1997}, the spatial tracer span should be approx 200 km or less, this points that the average spatial span of inhomogeneity tracers is less than the coronal separation. This further indicates the absence of significant turbulent plasma tracers and hence the weak cross-correlation between the two station data. Having said so, we do not imply a complete non-existence of such structures in the solar plasma. There is a possibility that these structures exist but they might decay so rapidly that their signatures are not recorded in the next station. Either way, in this experiment we couldn't detect any significant autocorrelation, which led to the conclusion that the cross-correlation technique in our present case was not able to yield significant solar wind velocities results, as done in the VLBI technique. It was in fact noticed during the entire period of our study and made us ponder that the radio-ray path is embedded in the region of solar wind where turbulence is still underdeveloped and in the source regime of turbulence spectrum.

Since the methodology based on cross-correlation of two station data posed some limitations in our case to use the VLBI technique for estimating the velocities in the inner corona, we decided to adopt another approach based on the isotropic quasi-static turbulence model which uses single station data to estimate the solar wind velocities.

\subsection{Isotropic Quasi-static Turbulence Model}

The dispersion relation of any plasma medium is the function of plasma frequency (that is the function of plasma densities) and the frequency of the incident electromagnetic wave \citep{FFChen1974}. For an electromagnetic wave passing through the coronal medium, as the turbulence is its inherent quantity at all heliolatitudes and heliocentric distances \citep{Coleman1968, Cranmer2007, Bruno2013}, the turbulent ionized medium induces dispersive effects in the phase and frequency of the traversing wave \citep{Tatarskii1971}. The impact of plasma density turbulence thus remains imprinted on the fluctuations in the phase of the emergent electromagnetic waves (corresponding to the Doppler shift in the frequency domain). These phase fluctuations thus can be analyzed to study the turbulence characteristics.
 
We considered one of the well-known models for studying solar wind turbulence - the isotropic three-dimensional quasi-static turbulence model as provided by the \citep{Armand1987, Chashei2007, Efimov2008, Efimov2010}. This model assumes a bulk outflow of `frozen-in' turbulence radially away from the coronal surface. It takes into account density inhomogeneities that follow a quasi-static isotropic turbulence pattern. The model was used to study the temporal frequency fluctuation turbulence spectra and extended further to estimate the plasma flow speeds perpendicular to the line of sight of the radio ray path using an approach as provided by \citep{Wexler2019}. Here it is assumed that the electron density inhomogeneities have three-dimensional turbulence spectra and are advected perpendicularly to the line of sight of the radio ray path with the background solar wind flow speeds.
 
The phase of EM waves of original transmitted frequency  $f_o$ are related to the phase of the received signal frequency $f_r$ by the rate of the change of relative spacecraft and receiver velocity $V_{rel}$ called as range rate Doppler along the line of sight (LOS) and rate of change refractive index, which is related to the electron density of the medium  \citep{Chashei2007, Wexler2019}. We based our analysis on the Frequency Fluctuation (FF) model of quasistatic isotropic turbulence as given in \citep{Wexler2019}, accordingly :

 \begin{equation}
 f_r = f_o - \frac{V_{rel}}{c}f_o + \frac{1}{2\pi} r_e \lambda \frac{d}{dt} \int^{L_{LOS}}_o n_e(s,t) ds 
\end{equation}

Here, $\lambda = c/f_o$ is the transmitted frequency wavelength, c is the speed of light, and $n_e$ is the electron number density in the solar corona along the line of sight length. Due to inherent turbulence in the medium, the density term $n_e$ includes a mean component and a fluctuating component of amplitude $ \delta n_e $ that contribute to the fluctuating component of FF, and ds is the vertical LOS integration path increment. $L_{LOS}$ is taken to be the magnetically determined correlation length, $r_e$ is the classical electron radius, $r_e = e^2 / (4 \pi \epsilon_o m_e c^2 ) $. If we remove the contribution arising due to the range rate Doppler (also mentioned as theoretical/geometrical),  the instantaneous change in FF is  $\delta f(t) = f_{obs}(t) - f_o$. Considering the density fluctuation term of the form $ \delta n_e(t) = \delta n_e exp^{-i \omega t}$, the time derivative relation of the FF becomes,

\begin{equation}
 \delta f(t) = \frac{1}{2 \pi} r_e \lambda L_{LOS} (-i \omega \delta n_e(t) )
\end{equation}

To analyze the power spectrum, we proceeded with the Fourier transform of the above equation. The power spectral density for a given data segment of temporal length T is denoted as $|FF(\omega)|^2 = (1/T) (\mathcal{F}\{\delta f(t)\}) (\mathcal{F}^*\{\delta f(t)\}) $,

\begin{equation}
 |FF(\omega)|^2  = \frac{1}{4\pi^2} r_e^2 \lambda^2 \omega^2 L_{LOS}^2 |\delta n_e(\omega)|^2 
\end{equation}

where $|\delta n_e (\omega)|^2$ is the corresponding power spectral density of electron concentration fluctuations,

This is the equation of density fluctuation of a single slab. We denoted FF in terms of oscillation frequency in Hz, $\nu = \omega / 2\pi$.  These FF are converted to radio wavelength normalized fluctuation measure FM ( by using $\sigma_{FM} = \sqrt{\sigma^2_{FF}}/\lambda $, based on rationale as given in \citep{Wexler2019}, as these fluctuations are sensitive in radio transmission wavelength $\lambda$. For interplanetary scintillation studies and coronal radio-sounding studies, it is typically considered that the strongest contribution to the frequency fluctuation comes from the electron plasma near the closest distance of the radio ray path from the solar surface along the LOS, which proximate point of radio ray path and is termed as 'solar offset SO' \citep{BirdEdenhofer1990, Ananthakrishnan1982} denoted by R. In terms of heliocentric distances and in the units of Solar radii ($R_\odot$) this radial distance will be noted as $R = rR_\odot$. The LOS integration path lengths are typically considered SO/2 in either direction of the proximate ray path point as it is believed that the coronal plasma electron density falls radially rapidly as we move away from the solar surface on either side of the solar proximate point. Hence, the effective total integration path length along LOS is generally taken equal to R $(=2*SO/2)$ in solar radio sounding experiments \citep{Wexler2019}. Here, it is assumed that the density fluctuations are aligned parallel to the magnetic field in a series of stacked slabs thus for the total number of stacks multiply by $R/L_{LOS}$ \citep{Wexler2019}. This renders, the relation between the Frequency fluctuation spectrum FM and the underlying density $n_e$ fluctuation spectrum as
 
 \begin{equation}
 |FM(\nu)|^2 = r_e^2 \nu^2 L_{LOS} R |\delta n_e (\nu)|^2 ,
\end{equation}

The frequency fluctuation FF values are obtained from the observed experimental parameters by processing the observed Doppler in the received signal as described in the previous section. This relation between FF power density is used to determine the electron power density fluctuations and the spectral indexes of the turbulence spectrum in the coronal medium. The fluctuation variance $\sigma^2_{FM} $ and $\sigma^2_{n_e} $ are defined for frequency integration range [a,b]  and are obtained by the numerical integration of the FF power spectrums,

\begin{equation}
 \sigma_{FM}^2 = \int^b_a |FM(\nu)|^2 d\nu
\end{equation}

\begin{equation}
 \sigma_{n_e}^2 = \int^b_a |\delta n_e(\nu)|^2 d\nu
\end{equation}

Equations 4, and 6 combined to give
\begin{equation}
 \sigma_{n_e}^2 = \frac{1}{(r_e^2 L_{LOS} R)} \int^b_a \frac{|FM(\nu)|^2}{\nu^2} d\nu 
\end{equation}

Now, To incorporate the effects of the turbulence in plasma, it is assumed that the FF power spectrum follows a power law of the form $\sigma_{FM}^2 = \zeta \nu^{-\beta} $, where exponent $\beta$ is the power index. After evaluating the integrals, the relations 5 and 7 are denoted as,

\begin{equation}
 \sigma_{FM}^2 = \frac{\zeta}{1-\beta} ({{\nu_b}^{(1-\beta)}}-{{\nu_a}^{(1-\beta)}}) 
\end{equation}

\begin{equation}
  \sigma_{n_e}^2 = \frac{-1}{1+\beta} \frac{\zeta}{r_e^2 L_{LOS} R} ({{\nu_b}^{(-1-\beta)}}-{{\nu_a}^{(-1-\beta)}})
\end{equation}

These expressions can be condensed as,
\begin{equation}
 \sigma_{FM}^2 = r_e^2 \nu_c^2 L_{LOS} R \sigma_{n_e}^2
\end{equation}

where the scaling frequency factor $\nu_c$ is defined as, 

\begin{equation}
 \nu_c^2 = \frac{\beta+1}{\beta-1} \frac{({{\nu_b}^{(1-\beta)}}-{{\nu_a}^{(1-\beta)}})}{({{\nu_b}^{(-1-\beta)}}-{{\nu_a}^{(-1-\beta)}})}
\end{equation}


\noindent hence, variance in density fluctuations can be estimated from variance in frequency fluctuations of the radio signal from the spectral index of the FF power spectrum.

Another useful quantity that is obtained from this analysis is the fractional density fluctuations $\epsilon$ that normalizes fluctuations with the mean local density $n_e$, and is defined as, 

\begin{equation}
\epsilon = \frac{\sigma_{n_e}}{n_e} = \frac{\sigma_{FM}}{r_e \nu_c n_e \sqrt{L_{LOS} R} }
\end{equation}

where $n_e(r)$ is the mean local electron density number. There are a number of coronal plasma expressions which are slight modifications of the Allen-Baumbech formula \citep{Patzold1987} that correlate the electron density with the radial distance from the solar surface. In this analysis, as our purpose has been to use the most appropriate relation that could fit electrons densities prevailing in the heliocentric distances of ($4-20 R_\odot$) and also during the minima phase of a solar cycle, we estimated it using the relation by \citep{Edenhofer1977} as given in \citep{Wexler2019}, which is based on ranging time-delays of the spacecraft radio signals,

\begin{equation}
 n_e(r) = (\frac{30}{r^6} + \frac{1}{r^{2.2}}) \times 10^{12}
\end{equation}

\noindent As their formula represented the number densities over $ 3<R<65R_\odot $ and also the calculations were done by Helios spacecraft in 1975, which was a minimum of solar cycle 21, thus this expression is well suited for our present study.

Thus this model utilizes the observational frequency fluctuations FF values to derive the FF turbulence power index $\beta$ and the fractional electron densities $\epsilon$. In order to derive these quantities we used the frequency residual values as recorded at the IDSN station Bangalore and then proceeded to find the power spectrum using the Fourier transform. The frequency fluctuation spectra were represented by a single power law in integration limits taken as $[\nu_{low}, \nu_{up}]$  where $\nu_{low} = 10^{-3} Hz$ is the length of the processed interval and $\nu_{up} = 5 \times 10^{-2} Hz$, beyond which system noise dominates and \citep{Efimov2008}. The power index $\beta$ was estimated as the slope of the curve when log values of frequency and PSD data were plotted and fitted with linear regression. The turbulence spectrum for different experiment days gave different values of $\beta$ which are tabulated in table \ref{data-table}. These $\beta$ values were used to calculate the scaling frequency factor $\nu_c$ as defined in eq (11). $\sigma_{FM}$ values were calculated by integrating FF power spectrum values in range $[\nu_{up}, \nu_{low}]$. We used the LOS element integration length as defined empirically in the \citep{Hollweg2010, Wexler2019} as a function of solar offset distances R $(R=rR_\odot)$  :

\begin{equation}
 L_{LOS} = 3.35 \times 10^6r^{0.918} 
\end{equation}

Using the above inputs for $\sigma_{FM}$, $\beta$, relations for $L_{LOS}$ ($meters$) and $n_e$ ($electrons/m^{-3}$) in the eq 12, we are able to estimate the fractional electron density fluctuation $\epsilon$ for the different experiments days. 

An intriguing interpretation is based on the slope of the power density spectrum which is related to the change in the turbulence regime of coronal plasma. In our study, we found out that for the entire range of probing heliocentric distances i.e. $5-13 R_\odot$ the power density spectrum has a flatter slope, which hints towards the fact that our radio ray path could be embedded in the region of weak turbulence of fast solar wind as discussed in the previous section. In the next section, we attempted to find the correlation between the slopes of these power density spectra and plasma flow speeds.

\subsection{Flow Speed Estimates}

The further goal is to connect the knowledge obtained from this FF model to the Efimov-Armand \citep{Armand1987, Efimov2008} turbulence model that takes into account the motion of irregularities embedded in the background solar wind, thus providing a way to estimate solar wind flow speeds.

This model is based on the theory of radio wave propagation in the turbulent medium \citep{Tatarskii1971}. It assumes a quasi-static isotropic 3D spatial electron density inhomogeneity spectrum. It was noted by \citep{Armand1987} that for plasma near the Sun, spatial fluctuations are assumed to be ``frozen-in'' in plasma and are convected outward at velocity $V_{sw}$ of the solar wind (Taylor hypothesis) across the line of sight of radio wave. The power index $\beta$ of frequency fluctuations density spectrum represents the turbulence spectrum. As seen in equation 5, integration of frequency spectral density ${|FM(\nu)|}^2$ gives the variance of frequency fluctuations $\sigma_{FM}^2$. The Efimov-Armand isotropic turbulence model relates these quantities as \citep{Armand1987, Efimov2008, Wexler2019} :

\begin{equation}
 \sigma_{FM}^2 = r_e^2 \biggl\{\frac{\beta}{ \pi (1-\beta)} (\nu_{up}^{1-\beta} - \nu_{low}^{1-\beta} ) \biggr \} \epsilon^2 n_e^2 L_e V_{sw}^{\beta+1} L_0^{-\beta}
\end{equation}

Where, $\nu_{up}$ and $\nu_{low}$ are upper and lower integration limits on the FF power spectrum, $L_o$ is the outer scale of turbulence and effective LOS integration path length $L_e \approx R $. For the outer scale of turbulence, we referred to the relation as given in \citep{Bird2002, Wexler2019} :
\begin{equation}
 L_o (r) = A_o r^{\mu}
\end{equation}

\noindent with $A_o = 0.23 \pm 0.11 R_\odot$ and $\mu = 0.82 \pm 0.13 $. Utilizing the values of other parameters as calculated from the above analysis we estimated the solar wind velocity values $V_{sw}$. The spectral index values $\beta$ were derived from the linear fit on the frequency fluctuation data, and the standard deviation on the fitted slope values gave us an estimate of the error in the calculated slope. This error was propagated to estimate the uncertainties in the estimated solar wind speed values that are presented in the table \ref{data-table}. We found that values for solar wind speed near 5 $R_\odot$ to 13 $R_\odot$ gradually increased from 250 km/s to 525 km/s. When compared with the general values as recorded from previous observations and models these values are a little higher \citep{Sheeley1997}.

\begin{table*}
 \centering
 \begin{tabular}{|c|c|c|c|c|c|c|c|}
 \hline
 Date & R($R_\odot$) & Heliolat & Power Index & $\sigma_{FM}$ & $\sigma_{ne}$ & $\epsilon$ & Flow Speeds \\
 & & (deg) & ($\beta\pm\delta\beta$) & $Hz$ & $m^{-3}$ & & (Km/s) \\
 \hline
 14-March-2021 & 12.82 & -24.67 & 0.490 $\pm$ 0.102 & 1.452 & 2.13 & 0.058 & 529.34 $\pm$ 18.82 \\
 \hline
 19-March-2021 & 8.71 & -37.66 & 0.572 $\pm$ 0.067 & 2.237 & 5.15 & 0.059 & 370.29 $\pm$ 2.475 \\
 \hline
 21-March-2021 & 7.24 & -46.78 & 0.298 $\pm$ 0.067 & 4.194 & 8.76 & 0.067 & 365.15 $\pm$ 39.38 \\
 \hline
 23-March-2021 & 6.09 & -58.57 & 0.249 $\pm$ 0.064 & 4.156 & 9.85 & 0.051 & 338.65 $\pm$ 47.16 \\
 \hline
 25-March-2021 & 5.28 & -75.75 & 0.417 $\pm$ 0.079 & 2.765 & 8.906 & 0.032 & 247.80 $\pm$ 10.92 \\
 \hline
 26-March-2021 & 5.09 & -86.02 & 0.450 $\pm$ 0.091 & 2.601 & 8.942 & 0.030 & 236.39 $\pm$ 9.01 \\
 \hline
 27-March-2021 & 5.08 & -82.65 & 0.409 $\pm$ 0.077 & 2.695 & 8.879 & 0.029 & 242.02 $\pm$ 10.96 \\
 \hline
 29-March-2021 & 5.57 & -62.23 & 0.362 $\pm$ 0.070 & 2.255 & 6.525 & 0.027 & 269.55 $\pm$ 16.66 \\
 \hline
 31-March-2021 & 6.62 & -46.54 & 0.564 $\pm$ 0.077 & 2.486 & 7.409 & 0.046 & 290.87 $\pm$ 2.29 \\
 \hline
 02-April-2021 & 7.98 & -35.72 & 0.249 $\pm$ 0.071 & 3.687 & 6.751 & 0.064 & 431.41 $\pm$ 67.80 \\
 \hline
 \end{tabular}
 \vspace*{.25cm}
 \caption{Results of Solar wind Speed estimates from Isotropic Quasi-static turbulence model. The power index value have been obtained from slope of power density spectrum of frequency fluctuation obtained from RO experiment signal recorded at IDSN station. Values of heliocentric distance $R_\odot$, heliolalitude of ray path proximate point corresponding to each experiment day is tabulated.}
 \label{data-table}
 
\end{table*}
 
  \begin{figure}
    \centering
    \resizebox{\hsize}{!}{\includegraphics{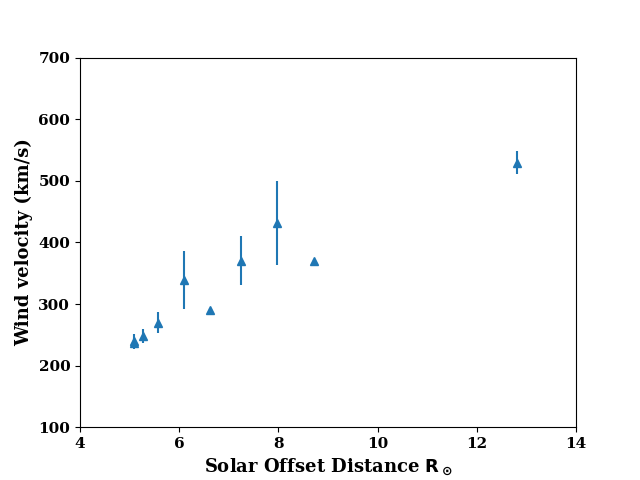}}
   
     \caption{Plasma flow velocities in the inner solar wind derived using quasi-static turbulence formulation from frequency residual data collected at IDSN station }
     \label{SolWinSpeed}%
     \end{figure}

Figure \ref{SolWinSpeed} shows the solar wind velocities as derived from the quasi-static turbulence model that tends to have values of around 250 Km/s near 5 $R_\odot$ and accelerates upto 550 Km/s near 13 $R_\odot$. Higher speed values indicate the fast solar wind streams which originate from the coronal holes that are usually found near the polar heliolatitudes \citep{Zirker1977}.
 
\subsection{Correlation with Solar EUV Images}

\begin{figure*}
     \centering
     \resizebox{\hsize}{!}{\includegraphics{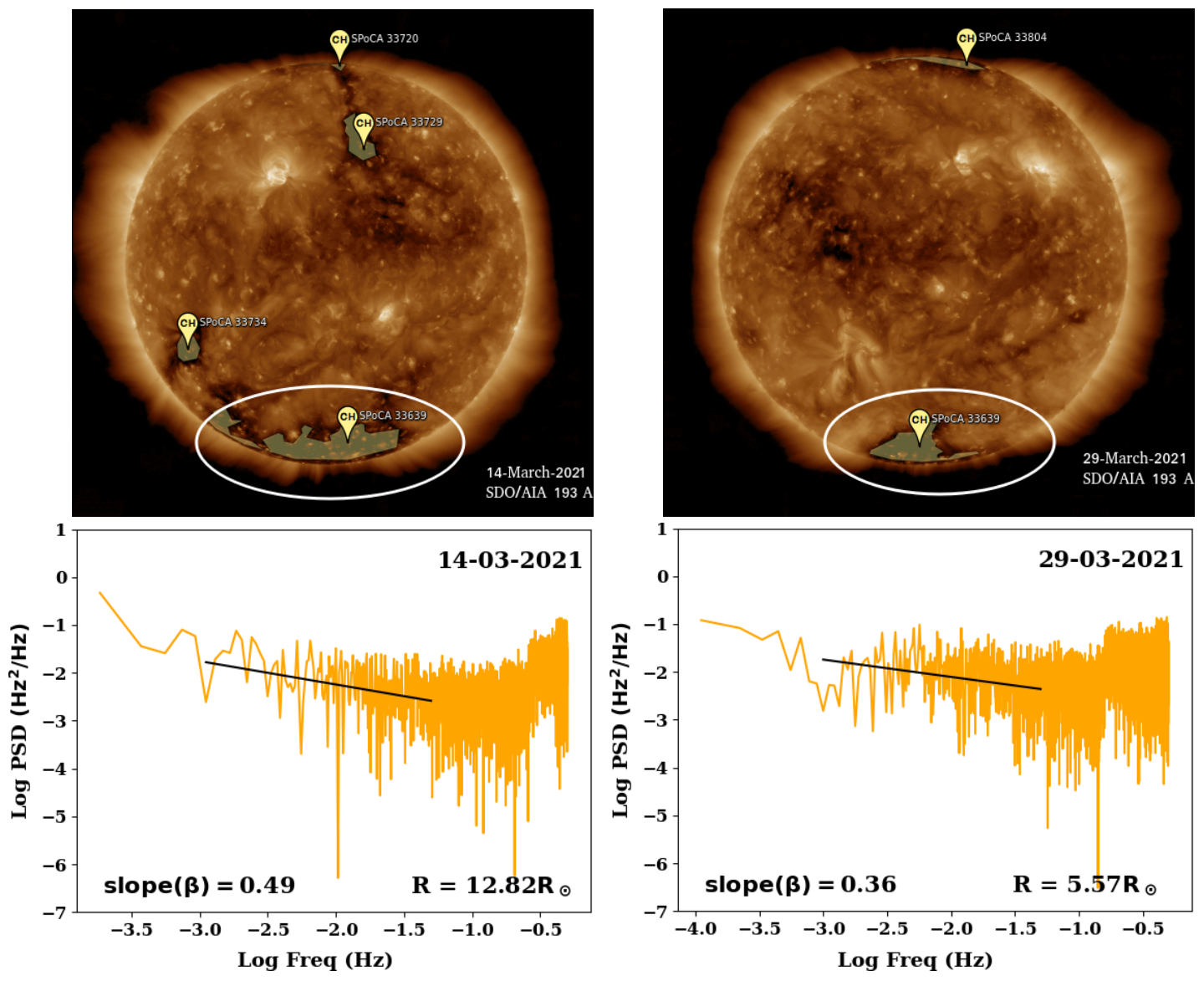}}
     \caption{ coronal Hole and FF power density Spectrum on the given experiment days. Upper panel shows the coronal holes imaged on solar surface by the SDO/AIA spacecraft. Lower panel shows the corrosponding power density spectrum of frequency fluctuations, heliocentric distances and slope. It can be observed that though at separate heliocentric distances the slope values are nearly identical, this is the consequence of the proximate point of our ray path being embeded in the region of influence of solar wind emanating from coronal holes.}
     \label{CHTurb}%
     \end{figure*}

In our study, we noticed that turbulence was flatter for most of the heliocentric distance as opposed to the expected increase in the slope values with the increasing heliocentric distances \citep{Jain2022}. The need for an explanation for these observed lower spectral index values combined with the higher average values of calculated solar wind speeds prompted us to consider looking at the coronal features prevailing on the solar surface during the experiment days. We referred to the SDO/AIA images in 193 $\AA$ wavelength for the given days and noticed the presence of a large coronal hole (christened as SPoCA 33639) near the southern heliocentric latitudes as shown in the Figure \ref{CHTurb}. The annotated coronal hole image is taken from the JHelioviewer interface \citep{Muller2017}. It existed in the same region where the proximate point of our radio ray path was embedded. The plausible explanation for the above observation becomes clearer when along with heliocentric distances we consider the heliolaltitudes of the proximate point of the ray path. Due to the relative inclination of Venus (and hence Akatsuki orbit) orbit with respect to Earth's ecliptic plane, the path of transmitted and received radio rays was mostly closer to the southern polar heliolatitudes (refer to Figure \ref{ROgeo}). The coronal hole corotated with the solar surface during the entire duration (both ingress and egress approx 20 days) of our experiment days, keeping the position of the proximate point always under the influence of high-speed plasma emerging from the coronal hole.

 Figure \ref{CHTurb} (upper panel) shows the presence of a coronal hole SPoCA 33639 in the SDO/AIA images during the observation period on the two observation days. The turbulence spectrum in Figure \ref{CHTurb} (lower panel) on those days corresponds to two different solar offset distances (on March 14, 2021, it was 12.80 $R_\odot$) and (on March 29, 2021, it was 5.57 $R_\odot$). The slope here represents the 'beta' index value, which corrosponds to the linear fit on the log-log curve of turbulence power spectrum ($=2*{\alpha_f}$ and defined in equation 8 of section 4.2). On both days, despite being at separate heliocentric distances, the slope is nearly identical (contrary to the expected steepness at larger distances). This interesting aspect of our observation is attributed to the presence of a coronal hole. The coronal holes are the region of open magnetic field lines which aids in the acceleration of the charged particles along them thus leading to the high-speed solar wind. The corroboration of EUV images further reinforces our observation of flatter turbulence spectrum and occurrence of high-speed flow as observed in the results of our solar wind speed models.

\section{Conclusions}

We used the coronal radio-sounding experiments, conducted in the inner heliosphere by the Akatsuki spacecraft, to study the characteristics of plasma turbulence and flow speeds. The radio signal obtained at the ground stations was spectrally analyzed to find the Doppler residual values. The turbulence spectrum showed the presence of an extended flatter source regime along the entire range of heliocentric distances. We estimated the flow speed of the turbulent plasma in the proximate ray path (the closest path of the radio waves to the solar surface) using the quasi-static turbulence model which is based on the estimation of flow velocities by assuming the propagation of radio waves in turbulent plasma medium where irregularities are assumed to be frozen-in and are advected with the background solar wind speeds. The flow speeds in the inner coronal regions between ($5-13 R_\odot$) are estimated as $220-550 Km/s$.  The SDO/AIA coronal images showed the presence of a large coronal hole, which explains the flatter slopes in the turbulence spectrum even with the increasing heliocentric distances and relatively higher flow speeds.

The inner solar coronal region is the main source of solar wind generation and acceleration. However, due to the constraints and paucity of in-situ measurements, it has remained a least explored region in the coronal studies. Recent observations from the Parker probe \citep{Kasper2021, Bandyopadhyay2022, Telloni2021, Zhao2022, Zhao2022b} and Solar Orbiter \citep{Antonucci2020, Adhikari2022} have provided promising insights into these studies, but remote sensing experiments which can transmit signals through these closer heliocentric distances are still one of the ingenious and convenient ways to explore the plasma properties and processes prevailing in this region.

The congruence of the flatter turbulence spectrum, higher flow speeds, and the presence of a coronal hole made this study interesting. This analysis aims at strengthening the observational evidence of turbulence regime and flow speeds present in the least probed inner coronal region.

\section*{Acknowledgments}

We acknowledge the efforts of the ground station team at IDSN, Bangalore India, and UDSC, Japan to track Akatsuki signals. Thanks are also due to Himanshu Pandey and other staff at ISSDC for the proper archival and dissemination of data for further analysis. Special thanks to Soumyaneal, Arya and Ajay of InSWIM lab for their help in error calculations. Anil Bhardwaj is supported by the Department of Space, India and he was also a JC Bose National Fellow during the period of this work.

\section*{Data Availability}
The Akatsuki solar radio occultation data are available in JAXA’s data archive \cite{Murakami2017} at \url{https://darts.isas.jaxa.jp/planet/project/akatsuki/rs.html.en}.



\bibliographystyle{mnras}
\bibliography{richa}

\begin{thebibliography}{}
\makeatletter
\relax
\def\mn@urlcharsother{\let\do\@makeother \do\$\do\&\do\#\do\^\do\_\do\%\do\~}
\def\mn@doi{\begingroup\mn@urlcharsother \@ifnextchar [ {\mn@doi@}
  {\mn@doi@[]}}
\def\mn@doi@[#1]#2{\def\@tempa{#1}\ifx\@tempa\@empty \href
  {http://dx.doi.org/#2} {doi:#2}\else \href {http://dx.doi.org/#2} {#1}\fi
  \endgroup}
\def\mn@eprint#1#2{\mn@eprint@#1:#2::\@nil}
\def\mn@eprint@arXiv#1{\href {http://arxiv.org/abs/#1} {{\tt arXiv:#1}}}
\def\mn@eprint@dblp#1{\href {http://dblp.uni-trier.de/rec/bibtex/#1.xml}
  {dblp:#1}}
\def\mn@eprint@#1:#2:#3:#4\@nil{\def\@tempa {#1}\def\@tempb {#2}\def\@tempc
  {#3}\ifx \@tempc \@empty \let \@tempc \@tempb \let \@tempb \@tempa \fi \ifx
  \@tempb \@empty \def\@tempb {arXiv}\fi \@ifundefined
  {mn@eprint@\@tempb}{\@tempb:\@tempc}{\expandafter \expandafter \csname
  mn@eprint@\@tempb\endcsname \expandafter{\@tempc}}}

\bibitem[\protect\citeauthoryear{Acton}{Acton}{1996}]{SPICEcite}
Acton C.,  1996, Planetary and Space Science, 44, 65

\bibitem[\protect\citeauthoryear{Adhikari, Zank  \& Zhao}{Adhikari
  et~al.}{2019}]{Adhikari2019}
Adhikari L.,  Zank G.~P.,   Zhao L.-L.,  2019, \mn@doi [The Astrophysical
  Journal] {10.3847/1538-4357/ab141c}, 876, 26

\bibitem[\protect\citeauthoryear{Adhikari, Zank, Telloni  \& Zhao}{Adhikari
  et~al.}{2022}]{Adhikari2022}
Adhikari L.,  Zank G.~P.,  Telloni D.,   Zhao L.-L.,  2022, \mn@doi [The
  Astrophysical Journal Letters] {10.3847/2041-8213/ac91c6}, 937, L29

\bibitem[\protect\citeauthoryear{{Alazraki} \& {Couturier}}{{Alazraki} \&
  {Couturier}}{1971}]{Alazraki1971}
{Alazraki} G.,  {Couturier} P.,  1971, \aap, \href
  {https://ui.adsabs.harvard.edu/abs/1971A&A....13..380A} {13, 380}

\bibitem[\protect\citeauthoryear{Ananthakrishnan \& Kaufman}{Ananthakrishnan \&
  Kaufman}{1982}]{Ananthakrishnan1982}
Ananthakrishnan S.,  Kaufman J.,  1982, Bulletin of the Astronomical Society of
  India, 10, 1

\bibitem[\protect\citeauthoryear{Antiochos, Miki{\'{c}}, Titov, Lionello  \&
  Linker}{Antiochos et~al.}{2011}]{Antiochos_2011}
Antiochos S.~K.,  Miki{\'{c}} Z.,  Titov V.~S.,  Lionello R.,   Linker J.~A.,
  2011, \mn@doi [The Astrophysical Journal] {10.1088/0004-637x/731/2/112}, 731,
  112

\bibitem[\protect\citeauthoryear{{Antonucci} et~al.,}{{Antonucci}
  et~al.}{2020}]{Antonucci2020}
{Antonucci} E.,  et~al., 2020, \mn@doi [\aap] {10.1051/0004-6361/201935338},
  \href {https://ui.adsabs.harvard.edu/abs/2020A&A...642A..10A} {642, A10}

\bibitem[\protect\citeauthoryear{{Armand}, {Efimov}  \& {Iakovlev}}{{Armand}
  et~al.}{1987}]{Armand1987}
{Armand} N.~A.,  {Efimov} A.~I.,   {Iakovlev} O.~I.,  1987, Astronomy and
  Astrophysics, 183, 135

\bibitem[\protect\citeauthoryear{Bandyopadhyay et~al.,}{Bandyopadhyay
  et~al.}{2022}]{Bandyopadhyay2022}
Bandyopadhyay R.,  et~al., 2022, \mn@doi [The Astrophysical Journal Letters]
  {10.3847/2041-8213/ac4a5c}, 926, L1

\bibitem[\protect\citeauthoryear{{Belcher}}{{Belcher}}{1971}]{Belcher1971}
{Belcher} J.~W.,  1971, \mn@doi [\apj] {10.1086/151105}, \href
  {https://ui.adsabs.harvard.edu/abs/1971ApJ...168..509B} {168, 509}

\bibitem[\protect\citeauthoryear{{Bird}}{{Bird}}{1982}]{Bird1982}
{Bird} M.~K.,  1982, \mn@doi [Space Sci.Rev.] {10.1007/BF00213250}, 33, 99

\bibitem[\protect\citeauthoryear{{Bird} \& {Edenhofer}}{{Bird} \&
  {Edenhofer}}{1990a}]{BirdEidenhofer1990}
{Bird} M.~K.,  {Edenhofer} P.,  1990a, in {Schwenn} R.,  {Marsch} E.,  eds, ,
  Physics of the Inner Heliosphere I.
p.~13, \mn@doi{10.1007/978-3-642-75361-9_2}

\bibitem[\protect\citeauthoryear{{Bird} \& {Edenhofer}}{{Bird} \&
  {Edenhofer}}{1990b}]{BirdEdenhofer1990}
{Bird} M.~K.,  {Edenhofer} P.,  1990b, in {Schwenn} R.,  {Marsch} E.,  eds, ,
  Physics of the Inner Heliosphere I.
p.~13, \mn@doi{10.1007/978-3-642-75361-9_2}

\bibitem[\protect\citeauthoryear{{Bird}, {Asmar}, {Brenkle}, {Edenhofer},
  {Paetzold}  \& {Volland}}{{Bird} et~al.}{1992}]{Bird1992}
{Bird} M.~K.,  {Asmar} S.~W.,  {Brenkle} J.~P.,  {Edenhofer} P.,  {Paetzold}
  M.,   {Volland} H.,  1992, \aaps, \href
  {https://ui.adsabs.harvard.edu/abs/1992A&AS...92..425B} {92, 425}

\bibitem[\protect\citeauthoryear{Bird, Efimov, Andreev, Samoznaev, Chashei,
  Edenhofer, Plettemeier  \& Wohlmuth}{Bird et~al.}{2002}]{Bird2002}
Bird M.,  Efimov A.,  Andreev V.,  Samoznaev L.,  Chashei I.,  Edenhofer P.,
  Plettemeier D.,   Wohlmuth R.,  2002, \mn@doi [Advances in Space Research]
  {https://doi.org/10.1016/S0273-1177(02)00334-4}, 30, 447

\bibitem[\protect\citeauthoryear{{Bruno} \& {Carbone}}{{Bruno} \&
  {Carbone}}{2013}]{Bruno2013}
{Bruno} R.,  {Carbone} V.,  2013, \mn@doi [Living Reviews in Solar Physics]
  {10.12942/lrsp-2013-2}, \href
  {https://ui.adsabs.harvard.edu/abs/2013LRSP...10....2B} {10, 2}

\bibitem[\protect\citeauthoryear{{Chandran} \& {Perez}}{{Chandran} \&
  {Perez}}{2019}]{Chandran2019}
{Chandran} B. D.~G.,  {Perez} J.~C.,  2019, \mn@doi [Journal of Plasma Physics]
  {10.1017/S0022377819000540}, \href
  {https://ui.adsabs.harvard.edu/abs/2019JPlPh..85d9009C} {85, 905850409}

\bibitem[\protect\citeauthoryear{{Chashei} \& {Shishov}}{{Chashei} \&
  {Shishov}}{1983}]{ChasheiModel1983}
{Chashei} I.~V.,  {Shishov} V.~I.,  1983, \sovast, 27, 346

\bibitem[\protect\citeauthoryear{Chashei~I.V, Plettemeier  \& Bird}{Chashei~I.V
  et~al.}{2005}]{Chashei2005}
Chashei~I.V A.I.~Efimov L.~S.,  Plettemeier D.,   Bird M.,  2005, Advances in
  Space Research, 36, 1454

\bibitem[\protect\citeauthoryear{Chashei, Efimov, Samoznaev, Bird  \&
  Pätzold}{Chashei et~al.}{2000}]{Chashei2000}
Chashei I.,  Efimov A.,  Samoznaev L.,  Bird M.,   Pätzold M.,  2000, \mn@doi
  [Advances in Space Research] {10.1016/S0273-1177(99)00614-6}, 25, 1973

\bibitem[\protect\citeauthoryear{{Chashei}, {Efimov}  \& {Bird}}{{Chashei}
  et~al.}{2007}]{Chashei2007}
{Chashei} I.~V.,  {Efimov} A.~I.,   {Bird} M.~K.,  2007, \mn@doi [Astronomical
  and Astrophysical Transactions] {10.1080/10556790701600515}, \href
  {https://ui.adsabs.harvard.edu/abs/2007A&AT...26..611C} {26, 611}

\bibitem[\protect\citeauthoryear{Chen}{Chen}{1974}]{FFChen1974}
Chen F.~F.,  1974, {Introduction to Plasma Physics}.
Springer US

\bibitem[\protect\citeauthoryear{{Chiba} et~al.,}{{Chiba}
  et~al.}{2022}]{Chiba2022}
{Chiba} S.,  et~al., 2022, \mn@doi [\solphys] {10.1007/s11207-022-01968-9},
  \href {https://ui.adsabs.harvard.edu/abs/2022SoPh..297...34C} {297, 34}

\bibitem[\protect\citeauthoryear{Choudhary, Ambili, Choudhury, Dhanya  \&
  Bhardwaj}{Choudhary et~al.}{2016}]{Choudhary2016}
Choudhary R.~K.,  Ambili K.~M.,  Choudhury S.,  Dhanya M.~B.,   Bhardwaj A.,
  2016, \mn@doi [Geophysical Research Letters]
  {https://doi.org/10.1002/2016GL070612}, 43, 10,025

\bibitem[\protect\citeauthoryear{{Coleman}}{{Coleman}}{1968}]{Coleman1968}
{Coleman} Paul~J. J.,  1968, \mn@doi [\apj] {10.1086/149674}, \href
  {https://ui.adsabs.harvard.edu/abs/1968ApJ...153..371C} {153, 371}

\bibitem[\protect\citeauthoryear{{Cranmer}, {van Ballegooijen}  \&
  {Edgar}}{{Cranmer} et~al.}{2007}]{Cranmer2007}
{Cranmer} S.~R.,  {van Ballegooijen} A.~A.,   {Edgar} R.~J.,  2007, \mn@doi
  [\apjs] {10.1086/518001}, \href
  {https://ui.adsabs.harvard.edu/abs/2007ApJS..171..520C} {171, 520}

\bibitem[\protect\citeauthoryear{Cranmer, Asgari-Targhi, Miralles, Raymond,
  Strachan, Tian  \& Woolsey}{Cranmer et~al.}{2015}]{Cranmer2015}
Cranmer S.~R.,  Asgari-Targhi M.,  Miralles M.~P.,  Raymond J.~C.,  Strachan
  L.,  Tian H.,   Woolsey L.~N.,  2015, \mn@doi [Philosophical Transactions of
  the Royal Society A: Mathematical, Physical and Engineering Sciences]
  {10.1098/rsta.2014.0148}, 373, 20140148

\bibitem[\protect\citeauthoryear{{Derby}}{{Derby}}{1978}]{Derby1978}
{Derby} N.~F. J.,  1978, \mn@doi [\apj] {10.1086/156451}, \href
  {https://ui.adsabs.harvard.edu/abs/1978ApJ...224.1013D} {224, 1013}

\bibitem[\protect\citeauthoryear{{Edenhofer}, {Lueneburg}, {Esposito},
  {Martin}, {Zygielbaum}, {Hansen}  \& {Hansen}}{{Edenhofer}
  et~al.}{1977}]{Edenhofer1977}
{Edenhofer} P.,  {Lueneburg} E.,  {Esposito} P.~B.,  {Martin} W.~L.,
  {Zygielbaum} A.~I.,  {Hansen} R.~T.,   {Hansen} S.~F.,  1977, Journal of
  Geophysics Zeitschrift Geophysik, \href
  {https://ui.adsabs.harvard.edu/abs/1977JGZG...42..673E} {42, 673}

\bibitem[\protect\citeauthoryear{{Efimov}, {Chashei}, {Bird}, {Plettemeier},
  {Edenhofer}, {Wohlmuth}, {Samoznaev}  \& {Lukanina}}{{Efimov}
  et~al.}{2005a}]{efimov2005}
{Efimov} A.~I.,  {Chashei} I.~V.,  {Bird} M.~K.,  {Plettemeier} D.,
  {Edenhofer} P.,  {Wohlmuth} R.,  {Samoznaev} L.~N.,   {Lukanina} L.~A.,
  2005a, \mn@doi [Advances in Space Research] {10.1016/j.asr.2005.02.030},
  \href {https://ui.adsabs.harvard.edu/abs/2005AdSpR..36.1448E} {36, 1448}

\bibitem[\protect\citeauthoryear{{Efimov}, {Chashei}, {Bird}, {Samoznaev}  \&
  {Plettemeier}}{{Efimov} et~al.}{2005b}]{efimov2005A}
{Efimov} A.~I.,  {Chashei} I.~V.,  {Bird} M.~K.,  {Samoznaev} L.~N.,
  {Plettemeier} D.,  2005b, \mn@doi [Astronomy Reports] {10.1134/1.1941491},
  49, 485

\bibitem[\protect\citeauthoryear{{Efimov}, {Samoznaev}, {Bird}, {Chashei}  \&
  {Plettemeier}}{{Efimov} et~al.}{2008}]{Efimov2008}
{Efimov} A.~I.,  {Samoznaev} L.~N.,  {Bird} M.~K.,  {Chashei} I.~V.,
  {Plettemeier} D.,  2008, \mn@doi [Advances in Space Research]
  {10.1016/j.asr.2008.03.025}, \href
  {https://ui.adsabs.harvard.edu/abs/2008AdSpR..42..117E} {42, 117}

\bibitem[\protect\citeauthoryear{{Efimov}, {Lukanina}, {Samoznaev}, {Chashei},
  {Bird}  \& {Plettemeier}}{{Efimov} et~al.}{2010}]{Efimov2010}
{Efimov} A.~I.,  {Lukanina} L.~A.,  {Samoznaev} L.~N.,  {Chashei} I.~V.,
  {Bird} M.~K.,   {Plettemeier} D.,  2010, \mn@doi [Astronomy Reports]
  {10.1134/S1063772910050082}, \href
  {https://ui.adsabs.harvard.edu/abs/2010ARep...54..446E} {54, 446}

\bibitem[\protect\citeauthoryear{{Geiss}, {Gloeckler}  \& {von
  Steiger}}{{Geiss} et~al.}{1995}]{Geiss1995}
{Geiss} J.,  {Gloeckler} G.,   {von Steiger} R.,  1995, \mn@doi [\ssr]
  {10.1007/BF00768753}, \href
  {https://ui.adsabs.harvard.edu/abs/1995SSRv...72...49G} {72, 49}

\bibitem[\protect\citeauthoryear{{Goldstein}}{{Goldstein}}{1978}]{Goldstein1978}
{Goldstein} M.~L.,  1978, \mn@doi [\apj] {10.1086/155829}, \href
  {https://ui.adsabs.harvard.edu/abs/1978ApJ...219..700G} {219, 700}

\bibitem[\protect\citeauthoryear{{Hewish}, {Scott}  \& {Wills}}{{Hewish}
  et~al.}{1964}]{Hewish1964}
{Hewish} A.,  {Scott} P.~F.,   {Wills} D.,  1964, \mn@doi [\nat]
  {10.1038/2031214a0}, \href
  {https://ui.adsabs.harvard.edu/abs/1964Natur.203.1214H} {203, 1214}

\bibitem[\protect\citeauthoryear{{Hollweg}}{{Hollweg}}{1986}]{Hollweg1986}
{Hollweg} J.~V.,  1986, \mn@doi [\jgr] {10.1029/JA091iA04p04111}, \href
  {https://ui.adsabs.harvard.edu/abs/1986JGR....91.4111H} {91, 4111}

\bibitem[\protect\citeauthoryear{{Hollweg}, {Cranmer}  \& {Chandran}}{{Hollweg}
  et~al.}{2010}]{Hollweg2010}
{Hollweg} J.~V.,  {Cranmer} S.~R.,   {Chandran} B. D.~G.,  2010, \mn@doi [\apj]
  {10.1088/0004-637X/722/2/1495}, \href
  {https://ui.adsabs.harvard.edu/abs/2010ApJ...722.1495H} {722, 1495}

\bibitem[\protect\citeauthoryear{{Imamura}, {Noguchi}, {Nabatov}, {Oyama},
  {Yamamoto}  \& {Tokumaru}}{{Imamura} et~al.}{2005}]{Imamura2005}
{Imamura} T.,  {Noguchi} K.,  {Nabatov} A.,  {Oyama} K.~I.,  {Yamamoto} Z.,
  {Tokumaru} M.,  2005, \mn@doi [\aap] {10.1051/0004-6361:20042614}, \href
  {https://ui.adsabs.harvard.edu/abs/2005A&A...439.1165I} {439, 1165}

\bibitem[\protect\citeauthoryear{Imamura et~al.,}{Imamura
  et~al.}{2011}]{Imamura2011}
Imamura T.,  et~al., 2011, Earth, planets and space, 63, 493

\bibitem[\protect\citeauthoryear{Imamura et~al.,}{Imamura
  et~al.}{2017}]{Imamura2017}
Imamura T.,  et~al., 2017, Earth, Planets and Space, 69, 1

\bibitem[\protect\citeauthoryear{Jain, Choudhary, Bhardwaj, Parikh, Dai  \&
  Roopa}{Jain et~al.}{2022}]{Jain2022}
Jain R.~N.,  Choudhary R.~K.,  Bhardwaj A.,  Parikh U.,  Dai B.~K.,   Roopa
  M.~V.,  2022, \mn@doi [Monthly Notices of the Royal Astronomical Society]
  {10.1093/mnras/stac056}, 511, 1750

\bibitem[\protect\citeauthoryear{Janardhan, Bird, Edenhofen, Wohlmuth,
  Plettemeier, Asmar, Patzold  \& Karl}{Janardhan et~al.}{1999}]{Janardhan1999}
Janardhan P.,  Bird M.~K.,  Edenhofen P.,  Wohlmuth R.,  Plettemeier D.,  Asmar
  S.~W.,  Patzold M.,   Karl J.,  1999, \mn@doi [Solar Physics]
  {10.1023/A:1005097028501}, 184, 157

\bibitem[\protect\citeauthoryear{Kasper et~al.,}{Kasper
  et~al.}{2021}]{Kasper2021}
Kasper J.~C.,  et~al., 2021, Phys. Rev. Lett., 127, 255101

\bibitem[\protect\citeauthoryear{Ma et~al.,}{Ma et~al.}{2021}]{MaoliMa2021}
Ma M.,  et~al., 2021, \mn@doi [The Astronomical Journal]
  {10.3847/1538-3881/ac0dc1}, 162, 141

\bibitem[\protect\citeauthoryear{Marsch}{Marsch}{1999}]{Marsch1999}
Marsch E.,  1999, Solar Wind Models from the Sun to 1 AU: Constraints by in
  Situ and Remote Sensing Measurements.
Springer Netherlands, Dordrecht, pp 1--24,
  \mn@doi{10.1007/978-94-015-9167-6_1}, \url
  {https://doi.org/10.1007/978-94-015-9167-6_1}

\bibitem[\protect\citeauthoryear{{Matthaeus}, {Zank}, {Oughton}, {Mullan}  \&
  {Dmitruk}}{{Matthaeus} et~al.}{1999}]{Matthaeus1999}
{Matthaeus} W.~H.,  {Zank} G.~P.,  {Oughton} S.,  {Mullan} D.~J.,   {Dmitruk}
  P.,  1999, \mn@doi [\apjl] {10.1086/312259}, \href
  {https://ui.adsabs.harvard.edu/abs/1999ApJ...523L..93M} {523, L93}

\bibitem[\protect\citeauthoryear{McComas et~al.,}{McComas
  et~al.}{2000}]{McComas2000}
McComas D.~J.,  et~al., 2000, \mn@doi [Journal of Geophysical Research: Space
  Physics] {https://doi.org/10.1029/1999JA000383}, 105, 10419

\bibitem[\protect\citeauthoryear{McComas, Elliott, Schwadron, Gosling, Skoug
  \& Goldstein}{McComas et~al.}{2003}]{McComas2003}
McComas D.~J.,  Elliott H.~A.,  Schwadron N.~A.,  Gosling J.~T.,  Skoug R.~M.,
   Goldstein B.~E.,  2003, \mn@doi [Geophysical Research Letters]
  {https://doi.org/10.1029/2003GL017136}, 30

\bibitem[\protect\citeauthoryear{{McIntosh}, {de Pontieu}, {Carlsson},
  {Hansteen}, {Boerner}  \& {Goossens}}{{McIntosh} et~al.}{2011}]{Mcintosh2011}
{McIntosh} S.~W.,  {de Pontieu} B.,  {Carlsson} M.,  {Hansteen} V.,  {Boerner}
  P.,   {Goossens} M.,  2011, \mn@doi [\nat] {10.1038/nature10235}, \href
  {https://ui.adsabs.harvard.edu/abs/2011Natur.475..477M} {475, 477}

\bibitem[\protect\citeauthoryear{{M{\"u}ller} et~al.,}{{M{\"u}ller}
  et~al.}{2017}]{Muller2017}
{M{\"u}ller} D.,  et~al., 2017, \mn@doi [\aap] {10.1051/0004-6361/201730893},
  \href {https://ui.adsabs.harvard.edu/abs/2017A&A...606A..10M} {606, A10}

\bibitem[\protect\citeauthoryear{Murakami, Ando, Imamura, Yamamoto  \&
  Hashimoto}{Murakami et~al.}{2017}]{Murakami2017}
Murakami S.,  Ando H.,  Imamura T.,  Yamamoto Y.,   Hashimoto G.~L.,  2017,
  \mn@doi [Institute of Space and Astronautical Science]
  {https://doi.org/10.17597/isas.darts/vco-00014}, p.~10

\bibitem[\protect\citeauthoryear{Nakamura et~al.,}{Nakamura
  et~al.}{2011}]{Nakamura2011}
Nakamura M.,  et~al., 2011, Earth, planets and space, 63, 443

\bibitem[\protect\citeauthoryear{{Parker}}{{Parker}}{1958}]{Parker1958}
{Parker} E.~N.,  1958, \mn@doi [\apj] {10.1086/146579}, \href
  {https://ui.adsabs.harvard.edu/abs/1958ApJ...128..664P} {128, 664}

\bibitem[\protect\citeauthoryear{{Patzold}, {Bird}, {Volland}, {Levy}, {Seidel}
   \& {Stelzried}}{{Patzold} et~al.}{1987}]{Patzold1987}
{Patzold} M.,  {Bird} M.~K.,  {Volland} H.,  {Levy} G.~S.,  {Seidel} B.~L.,
  {Stelzried} C.~T.,  1987, \mn@doi [\solphys] {10.1007/BF00167401}, \href
  {https://ui.adsabs.harvard.edu/abs/1987SoPh..109...91P} {109, 91}

\bibitem[\protect\citeauthoryear{{P{\"a}tzold} et~al.,}{{P{\"a}tzold}
  et~al.}{2004}]{Patzold2004}
{P{\"a}tzold} M.,  et~al., 2004, in {Wilson} A.,  {Chicarro} A.,  eds,  ESA
  Special Publication Vol. 1240, Mars Express: the Scientific Payload. pp
  141--163

\bibitem[\protect\citeauthoryear{{Pätzold}, {Karl}  \& {Bird}}{{Pätzold}
  et~al.}{1996}]{Patzold1996}
{Pätzold} M.,  {Karl} J.,   {Bird} M.~K.,  1996, \aap, \href
  {https://ui.adsabs.harvard.edu/abs/1996A&A...316..449P} {316, 449}

\bibitem[\protect\citeauthoryear{Pätzold, {Tsurutani}  \& {Bird}}{Pätzold
  et~al.}{1997}]{patzold1997}
Pätzold M.,  {Tsurutani} B.~T.,   {Bird} M.~K.,  1997, \mn@doi [\jgr]
  {10.1029/97JA01868}, \href
  {https://ui.adsabs.harvard.edu/abs/1997JGR...10224151P} {102, 24151}

\bibitem[\protect\citeauthoryear{Pätzold, Hahn, Tellmann, Haeusler, Bird,
  Tyler, Asmar  \& Tsurutani}{Pätzold et~al.}{2012}]{Patzold2012}
Pätzold M.,  Hahn M.,  Tellmann S.,  Haeusler B.,  Bird M.,  Tyler G.,  Asmar
  S.,   Tsurutani B.,  2012, \mn@doi [Solar Physics]
  {10.1007/s11207-012-9991-y}, 279

\bibitem[\protect\citeauthoryear{Roberts, Goldstein, Matthaeus  \&
  Ghosh}{Roberts et~al.}{1992}]{Roberts1992}
Roberts D.,  Goldstein M.,  Matthaeus W.,   Ghosh S.,  1992, \mn@doi [Journal
  of Geophysical Research] {10.1029/92JA01144}, 97

\bibitem[\protect\citeauthoryear{{Sakurai}}{{Sakurai}}{2017}]{Sakurai2017}
{Sakurai} T.,  2017, \mn@doi [Proceedings of the Japan Academy, Series B]
  {10.2183/pjab.93.006}, \href
  {https://ui.adsabs.harvard.edu/abs/2017PJAB...93...87S} {93, 87}

\bibitem[\protect\citeauthoryear{{Sheeley} et~al.,}{{Sheeley}
  et~al.}{1997}]{Sheeley1997}
{Sheeley} N.~R.,  et~al., 1997, \mn@doi [\apj] {10.1086/304338}, \href
  {https://ui.adsabs.harvard.edu/abs/1997ApJ...484..472S} {484, 472}

\bibitem[\protect\citeauthoryear{{Shoda}, {Suzuki}, {Asgari-Targhi}  \&
  {Yokoyama}}{{Shoda} et~al.}{2019}]{Shoda2019}
{Shoda} M.,  {Suzuki} T.~K.,  {Asgari-Targhi} M.,   {Yokoyama} T.,  2019,
  \mn@doi [\apjl] {10.3847/2041-8213/ab2b45}, \href
  {https://ui.adsabs.harvard.edu/abs/2019ApJ...880L...2S} {880, L2}

\bibitem[\protect\citeauthoryear{{Spangler}, {Kavars}, {Kortenkamp}, {Bondi},
  {Mantovani}  \& {Alef}}{{Spangler} et~al.}{2002}]{Spangler2002}
{Spangler} S.~R.,  {Kavars} D.~W.,  {Kortenkamp} P.~S.,  {Bondi} M.,
  {Mantovani} F.,   {Alef} W.,  2002, \mn@doi [\aap]
  {10.1051/0004-6361:20020028}, \href
  {https://ui.adsabs.harvard.edu/abs/2002A&A...384..654S} {384, 654}

\bibitem[\protect\citeauthoryear{Tatarski}{Tatarski}{1971}]{Tatarskii1971}
Tatarski V.,  1971, ~The{\oe} Effects of the Turbulent Atmosphere on Wave
  Propagation.
Israel Programm for Scientific Translations, \url
  {https://books.google.co.in/books?id=eN1AvwEACAAJ}

\bibitem[\protect\citeauthoryear{Telloni et~al.,}{Telloni
  et~al.}{2021}]{Telloni2021}
Telloni D.,  et~al., 2021, \mn@doi [The Astrophysical Journal Letters]
  {10.3847/2041-8213/ac282f}, 920, L14

\bibitem[\protect\citeauthoryear{Tripathi \& Choudhary}{Tripathi \&
  Choudhary}{2022}]{Tripathi-Choudhary2022}
Tripathi K.~R.,  Choudhary R.~K.,  2022, Earth and Space Science, 9,
  e2022EA002326

\bibitem[\protect\citeauthoryear{Tripathi, Choudhary  \& Jayalal}{Tripathi
  et~al.}{2022}]{Tripathi2022}
Tripathi K.~R.,  Choudhary R.~K.,   Jayalal L.,  2022, Mon. Not. R. Astron.
  Soc., 517, 776

\bibitem[\protect\citeauthoryear{Upendran, Tripathi, Mithun, Vadawale  \&
  Bhardwaj}{Upendran et~al.}{2022}]{Upendran2022}
Upendran V.,  Tripathi D.,  Mithun N. P.~S.,  Vadawale S.,   Bhardwaj A.,
  2022, \mn@doi [The Astrophysical Journal Letters] {10.3847/2041-8213/aca078},
  940, L38

\bibitem[\protect\citeauthoryear{Vadawale et~al.,}{Vadawale
  et~al.}{2021}]{Vadawale2021}
Vadawale S.~V.,  et~al., 2021, \mn@doi [The Astrophysical Journal Letters]
  {10.3847/2041-8213/abf0b0}, 912, L13

\bibitem[\protect\citeauthoryear{Verdini \& Velli}{Verdini \&
  Velli}{2007}]{Verdini2007}
Verdini A.,  Velli M.,  2007, \mn@doi [The Astrophysical Journal]
  {10.1086/510710}, 662, 669

\bibitem[\protect\citeauthoryear{Wexler, Hollweg, Efimov, Song, Jensen,
  Lionello, Vierinen  \& Coster}{Wexler et~al.}{2019a}]{Wexler2019b}
Wexler D.~B.,  Hollweg J.~V.,  Efimov A.~I.,  Song P.,  Jensen E.~A.,  Lionello
  R.,  Vierinen J.,   Coster A.~J.,  2019a, \mn@doi [Journal of Geophysical
  Research: Space Physics] {https://doi.org/10.1029/2019JA026937}, 124, 7761

\bibitem[\protect\citeauthoryear{Wexler, Hollweg, Efimov, Lukanina, Coster,
  Vierinen  \& Jensen}{Wexler et~al.}{2019b}]{Wexler2019}
Wexler D.~B.,  Hollweg J.~V.,  Efimov A.~I.,  Lukanina L.~A.,  Coster A.~J.,
  Vierinen J.,   Jensen E.~A.,  2019b, \mn@doi [The Astrophysical Journal]
  {10.3847/1538-4357/aaf6a8}, 871, 202

\bibitem[\protect\citeauthoryear{{Wohlmuth}, {Plettemeier}, {Edenhofer},
  {Bird}, {Efimov}, {Andreev}, {Samoznaev}  \& {Chashei}}{{Wohlmuth}
  et~al.}{2001}]{Wohlmuth2001}
{Wohlmuth} R.,  {Plettemeier} D.,  {Edenhofer} P.,  {Bird} M.~K.,  {Efimov}
  A.~I.,  {Andreev} V.~E.,  {Samoznaev} L.~N.,   {Chashei} I.~V.,  2001,
  \mn@doi [Space Sci.Rev.2008] {10.1023/A:1011845221808}, 97, 9

\bibitem[\protect\citeauthoryear{Woo}{Woo}{1995}]{Woo1995}
Woo R.,  1995, \mn@doi [Geophysical Research Letters]
  {https://doi.org/10.1029/95GL01315}, 22, 1393

\bibitem[\protect\citeauthoryear{Woo \& Armstrong}{Woo \&
  Armstrong}{1979}]{WooArmstrong1979}
Woo R.,  Armstrong J.~W.,  1979, \mn@doi [Journal of Geophysical Research:
  Space Physics] {https://doi.org/10.1029/JA084iA12p07288}, 84, 7288

\bibitem[\protect\citeauthoryear{{Woo}, {Yang}, {Yip}  \& {Kendall}}{{Woo}
  et~al.}{1976}]{Woo1976}
{Woo} R.,  {Yang} F.~C.,  {Yip} K.~W.,   {Kendall} W.~B.,  1976, \mn@doi [\apj]
  {10.1086/154861}, \href
  {https://ui.adsabs.harvard.edu/abs/1976ApJ...210..568W} {210, 568}

\bibitem[\protect\citeauthoryear{{Yakovlev}, {Efimov}, {Razmanov}  \&
  {Shtrykov}}{{Yakovlev} et~al.}{1980}]{Yakovlev1980}
{Yakovlev} O.~I.,  {Efimov} A.~I.,  {Razmanov} V.~M.,   {Shtrykov} V.~K.,
  1980, \sovast, \href {https://ui.adsabs.harvard.edu/abs/1980SvA....24..454Y}
  {24, 454}

\bibitem[\protect\citeauthoryear{Zhao, Zank, Telloni, Stevens, Kasper  \&
  Bale}{Zhao et~al.}{2022a}]{Zhao2022b}
Zhao L.-L.,  Zank G.~P.,  Telloni D.,  Stevens M.,  Kasper J.~C.,   Bale S.~D.,
   2022a, \mn@doi [The Astrophysical Journal Letters]
  {10.3847/2041-8213/ac5fb0}, 928, L15

\bibitem[\protect\citeauthoryear{Zhao, Zank, Adhikari, Telloni, Stevens,
  Kasper, Bale  \& Raouafi}{Zhao et~al.}{2022b}]{Zhao2022}
Zhao L.-L.,  Zank G.~P.,  Adhikari L.,  Telloni D.,  Stevens M.,  Kasper J.~C.,
   Bale S.~D.,   Raouafi N.~E.,  2022b, \mn@doi [The Astrophysical Journal
  Letters] {10.3847/2041-8213/ac8353}, 934, L36

\bibitem[\protect\citeauthoryear{Zirker}{Zirker}{1977}]{Zirker1977}
Zirker J.~B.,  1977, \mn@doi [Reviews of Geophysics]
  {https://doi.org/10.1029/RG015i003p00257}, 15, 257

\makeatother
\end{thebibliography}

\bsp	
\label{lastpage}
\end{document}